\documentclass{ws-ijmpd}
\usepackage[super,compress]{cite}
\usepackage{bbold}
\begin{document}

\markboth{S.F.~Hassan, Angnis~Schmidt-May, Mikael~von~Strauss}
{Particular Solutions in Bimetric Theory and Their Implications}

%
\catchline{}{}{}{}{}
%
%
\newcommand{\ba}{{\begin{align}}}
\newcommand{\ea}{{\end{align}}}
\newcommand{\be}{\begin{eqnarray}}
\newcommand{\ee}{\end{eqnarray}}
\newcommand{\beqn}{\begin{eqnarray}}
\newcommand{\eeqn}{\end{eqnarray}}
\newcommand{\dd}{\mathrm{d}}
\newcommand{\nn}{\nonumber}
\newcommand{\Tr}{\mathrm{Tr}}
\newcommand{\mm}{\mathrm{m}}
\newcommand{\gmn}{g_{\mu\nu}}
\newcommand{\fmn}{f_{\mu\nu}}
\newcommand{\bgmn}{\bar{g}_{\mu\nu}}
\newcommand{\bfmn}{\bar{f}_{\mu\nu}}
\newcommand{\emn}{\eta_{\mu\nu}}
\newcommand{\G}{\mathcal{G}}
\newcommand{\ha}{\hat\alpha}
\newcommand{\Pmn}{P^\mu_{\ph\mu\nu}}
\newcommand{\Smn}{S^\mu_{\ph\mu\nu}}
\newcommand{\pmn}{P_{\mu\nu}}
\newcommand{\gmnp}{g'_{\mu\nu}}
\newcommand{\fmnp}{f'_{\mu\nu}}
\newcommand{\p}{\partial}
\def\ph{\phantom}
%
\title{Particular Solutions in Bimetric Theory and Their Implications}

\author{S.F.~Hassan}

\address{Department of Physics \& 
        The Oskar Klein Centre,\\
        Stockholm University, AlbaNova University Centre, 
        SE-106 91 Stockholm, Sweden\\
fawad@fysik.su.se}

\author{Angnis~Schmidt-May}

\address{Department of Physics \& 
        The Oskar Klein Centre,\\
        Stockholm University, AlbaNova University Centre, 
        SE-106 91 Stockholm, Sweden\\
angnis.schmidt-may@fysik.su.se}

\author{Mikael~von~Strauss}

\address{UPMC-CNRS, UMR7095,
 Institut d'Astrophysique de Paris, GReCO,\\
 98bis boulevard Arago, F-75014 Paris, France\\
strauss@iap.fr}

\maketitle

\begin{history}
\received{Day Month Year}
\revised{Day Month Year}
\end{history}

\begin{abstract}
Ghost-free bimetric theory can describe gravity in the
  presence of an extra spin-2 field. We study certain aspects of
  dynamics in this theory: (i) It is shown that if either of the
  metrics is an Einstein solution then the other is always forced to
  be Einstein, too. For a class of bimetric models this constraint is
  stronger and as soon as one metric is Einstein, the other metric is
  forced to be proportional to it. As a consequence, the models in this
  class avoid a branch of pathological solutions that exhibit
  determinant singularities or nonlinear ghosts. These constraints
  persists in a generalized form when sources are included, but are
  destroyed in the massive gravity limit of the theory. (ii) For
  another class of bimetric models, we show the existence of solutions
  that do not admit a massive gravity limit. A bimetric model that
  could exhibit a nonlinear version of ``partially massless'' symmetry
  belongs to both these classes. It is argued that if such a model 
  exist, its symmetry will not survive in the massive gravity limit.
\end{abstract}


\ccode{PACS numbers: 04.50.Kd}


\section{Introduction \& Results}
The ghost-free bimetric theory \cite{Hassan:2011zd, Hassan:2011ea}
describes non-derivative interactions of two spin-2 fields, say,
$\gmn$ and $\fmn$, through a potential $V(g,f,\beta_n)$. The
interactions are parametrized by five parameters $\beta_n$ ($n=0\cdots
4$). This theory can be interpreted as a theory of gravity, with a
metric $\gmn$, in the presence of an extra spin-2 field $\fmn$. In one
limit, $\gmn$ is mostly massless and the theory is closer to General
Relativity \cite{Hassan:2012wr}. In an opposite limit, the field
$\fmn$ gets frozen to a fixed background and $\gmn$ becomes a massive
spin-2 field. This is the massive gravity limit~\cite{deRham:2010ik,
  deRham:2010kj, Hassan:2011vm, Hassan:2011hr, Hassan:2011tf}. The
crucial point is that these theories do not contain the Boulware-Deser
ghost~\cite{Boulware:1972zf}. However, the implications of such spin-2
interactions and many aspects of their dynamics are far from fully
understood.

To be viable, bimetric models must admit solutions without
pathologies. Another interesting aspect is the possible relevance to
nonlinear realizations of partial masslessness. In this paper we study
some aspects of dynamics of bimetric models that are relevant for
these issues. Most of our results are in the context of two classes of
bimetric models, that we call the ``$\beta_i$-models'' and the
``symmetric'' models.

In the $\beta_i$-models, $\beta_i$ denotes one of the three parameters
$\beta_1, \beta_2, \beta_3$, while the other two are set to zero. The
remaining parameters $\beta_0$ and $\beta_4$ take generic values. Thus,
for example, the $\beta_1$ model is defined by $\beta_2= \beta_3=0$.
We show that the dynamics of these models are more constrained than in
generic bimetric models. In particular they avoid pathological
branches of solutions that arise along with healthy solutions.

In the symmetric models, the parameters $\beta_n$ satisfy the property
$\alpha^{4-n}\beta_n=\alpha^n\beta_{4-n}$, where $\alpha=m_f/m_g$ is
the ratio of the Planck masses of the two spin-2 fields. We show that
this class of models always contains solutions that do not admit a
massive gravity limit. This result is relevant for the viability of
the proposal that the bimetric framework could admit a nonlinear
generalization of linear partial masslessness. 

All our considerations here are at the classical level and in this work we do not address the question whether quantum corrections maintain the structure of the above models or not. Our results are
summarized below.

\subsection{Summary of results}
\paragraph{Einstein and proportional solutions:}
In the absence of sources, bimetric equations always admit solutions
where at least one metric is Einstein (that is, a solution to
Einstein's vacuum equations).
\begin{itemize} \itemsep0em  
\item We show that in any bimetric theory, if one of the metrics is
  Einstein, then the dynamics also force the other metric to be
  Einstein. 
\item In the $\beta_i$-models (defined above) the constraint is
  stronger: If either metric is Einstein, the other metric is not only
  Einstein, but also proportional to the first metric.  
\item Similar statements apply to bimetric solutions in the presence
  of matter sources, but now only when the matter couplings allow for
  GR solutions (that is, solutions of Einstein's equations in the
  presence of sources).
\item The $\beta_i$ models do not contain branches of ill-behaved
  solutions exhibiting determinant singularities \cite{Gratia:2013uza}
  or nonlinear ghosts \cite{Gumrukcuoglu:2011zh,
    DeFelice:2012mx}. Such solutions, that can arise in other models,
  represent examples of non-proportional Einstein backgrounds not
  permitted in the $\beta_i$-models.
\end{itemize} 
It is obvious that the above constraints do not arise in the massive
gravity limit of the theory, where the reference metric is always
Einstein without forcing the massive metric to be so.

\paragraph{Symmetric models and solutions with no massive gravity
  limit:} Massive gravity is defined with a fixed reference metric
$\bfmn$. Hence the massive gravity limit of bimetric theory also
involves truncating its solution space to a class that leads to the
specific $\bfmn$, and not to other reference metrics. Obviously, only
those properties of bimetric theory that admit a similar truncation
have a chance of surviving the massive gravity limit. There also exist
bimetric solutions that lead to singular metrics and have no massive
gravity limit. 
\begin{itemize} \itemsep 0em   
\item We show that symmetric bimetric models, with $\alpha^{4-n}
  \beta_n=\alpha^n\beta_{4-n}$, always contain solutions with no
  massive gravity limit, giving rise to singular metrics in the limit.  
\item The symmetric models include the candidate PM bimetric
  model (which is also a $\beta_2$-model). Based on known results, we
  argue that the PM gauge symmetry requires the presence of solutions
  with no massive gravity limit and that such a gauge symmetry, if
  present, will not admit a non-singular massive gravity truncation,
  except for Einstein backgrounds (and some other simple solutions).
\end{itemize}
From this point of view, the absence of a nonlinear PM symmetry in 
massive gravity is expected and does not rule out a nonlinear
realization in the PM bimetric model or some generalization of it. 

The paper is organized as follows. In section \ref{sec:intro} we
review some aspects of bimetric theory required for our analysis and
make the notion of its massive gravity limit
precise. Section~\ref{sec:conseq} studies Einstein backgrounds and
demonstrates the constraining nature of the bimetric equations of
motion. We also point out the absence of known ill-behaved solutions
in $\beta_i$-models.  In section \ref{sec:behav}, we analyze solutions
with no massive gravity limit and discuss the implications for some
counter arguments against the candidate PM bimetric model.

\section{Review of bimetric theory and its massive gravity limit}
\label{sec:intro}
Here we review some aspects of the ghost-free bimetric theory needed
for the rest of the paper. 
\subsection{Action and equations of motion}
The ghost-free bimetric action for two symmetric rank-two tensors
$\gmn$ and $\fmn$ is given by~\cite{Hassan:2011zd}, 
\beqn
\label{act}
S[g,f]=\int\dd^4 x\left(m_g^2~\sqrt{|g|}~R(g)+m_f^2~\sqrt{|f|}~R(f)
-2m^4~\sqrt{|g|}~V(S)\right)\,. 
\eeqn
Here, the kinetic structures for both tensors are the ordinary
Einstein-Hilbert terms with Planck masses $m_g$ and $m_f$. The
interaction potential $V$ is a linear combination of elementary
symmetric polynomials $e_n(S)$ constructed out of the square-root 
matrix $S\equiv\sqrt{g^{-1}f}$, 
\beqn
\label{vdef}
V(S)= \sum_{n=0}^{4}\beta_ne_n(S)\,,
\eeqn
with arbitrary coefficients $\beta_n$. The $e_n(S)$ are given by the
recursive relations,  
\beqn
e_n(S)=\tfrac{1}{n}\sum_{k=0}^{n-1}(-1)^{k+n+1}\mathrm{Tr}(S^{n-k})
e_k(S)\,,\qquad e_0(S)=1\,.
\eeqn
So that $e_4(S)=\det S$ and $e_n(S)=0$ for $n>4$, if $S$ is a
$(4\times4)$-matrix. $V(S)$ generalizes the massive gravity potential
\cite{deRham:2010kj} as formulated in \cite{Hassan:2011vm} and shown
to be free of the Boulware-Deser ghost instability
\cite{Hassan:2011hr,Hassan:2011tf,Hassan:2011zd,Hassan:2011ea}.   
In this formulation it is evident that the relation
$\sqrt{|g|}~e_{n}(S)=\sqrt{|f|}~e_{4-n}(S^{-1})$ implies the property 
\cite{Hassan:2011zd}, 
\beqn
\sqrt{|g|}~V\left(\sqrt{g^{-1}f}\,;\beta_n\right)=
\sqrt{|f|}~V\left(\sqrt{f^{-1}g}\,;\beta_{4-n}\right)\,. 
\label{eid}
\eeqn
Variation of (\ref{act}) with respect to $g$ and $f$ gives two sets of
equations of motion. In terms of the Einstein tensor
$\G_{\mu\nu}=R_{\mu\nu}-\frac{1}{2}\gmn R$ for $\gmn$ (and similarly
$\tilde\G_{\mu\nu}$ for $\fmn$), they read, 
\beqn
\G_{\mu\nu}+\tfrac{m^4}{m_g^2}V_{\mu\nu}= 0\,,
\qquad
\alpha^2\tilde{\G}_{\mu\nu}+\tfrac{m^4}{m_g^2}\tilde V_{\mu\nu}=0\,, 
\label{gf-eq} 
\eeqn
where, 
\beqn
V_{\mu\nu}\equiv-\frac{2}{\sqrt{|g|}}\frac{\p (\sqrt{|g|}~V)}
{\p g^{\mu\nu}}\,,\qquad 
\tilde{V}_{\mu\nu}\equiv-\frac{2}{\sqrt{|f|}}\frac{\p
(\sqrt{|g|}~V)}{\p f^{\mu\nu}}\,.
\label{VsDef}
\eeqn
In terms of matrices $Y_{(n)}(S)$, the potential contributions have
the structure \cite{Hassan:2011vm,Hassan:2012rq},   
\beqn
V_{\mu\nu}\equiv g_{\mu\rho}\sum_{n=0}^3 \beta_n[Y_{(n)}(S)]^\rho_{~\nu}
\,,\qquad
\tilde{V}_{\mu\nu}\equiv f_{\mu\rho} \sum_{n=0}^3 \beta_{4-n}
[Y_{(n)}(S^{-1})]^\rho_{~\nu}\,,
\eeqn
where $Y_{(n)}$ are given as functions of the matrix $S$ by,
\beqn
[Y_{(n)}(S)]^\rho_{~\nu}\equiv\sum_{k=0}^n(-1)^ke_k(S)\,[S^{n-k}]^\rho_{~\nu}\,.   
\eeqn
$\tilde V_{\mu\nu}$ can be directly obtained from $V_{\mu\nu}$ through
$S\rightarrow S^{-1}$, $\beta_n\rightarrow \beta_{4-n}$ as implied by 
\eqref{eid}. For later use we note an identity for $Y_{(n)}$,  
\beqn
\Tr(S \,Y_{(n)}(S))&=&(-1)^n(n+1)e_{n+1}(S)\,. \label{Yid}
\eeqn 
The $\gmn$ equations and the Bianchi identity, $\nabla^\sigma
\G_{\sigma\nu}=0$, give the Bianchi constraint,
\beqn
\label{bianchi} 
\nabla^\sigma V_{\sigma\nu}=0\,.  
\eeqn 
There is a similar constraint arising from the $\fmn$ equations but
due to the overall diffeomorphism invariance the two constraints are
equivalent. 

Note that due to the property \eqref{eid}, the bimetric action
(\ref{act}) retains its form under the simultaneous interchanges,
\beqn
\label{intsym1}
\gmn\leftrightarrow \fmn\,,
\qquad m_g\leftrightarrow m_f\,,
\qquad \beta_n\rightarrow \beta_{4-n}\,,
\eeqn 
or the related transformations,
\beqn
\label{intsym}
\alpha^{-1}\gmn\leftrightarrow \alpha\fmn\,,
\qquad \alpha^{4-n}\beta_n\rightarrow \alpha^{n}\beta_{4-n}
\,,\qquad\text{where}~~\alpha\equiv m_f/m_g\,.
\eeqn
We emphasize that \eqref{intsym} keeps $m_g$ and $m_f$ unaltered and
so is different from, and less restrictive than,
\eqref{intsym1}. These transformations map the action $S[g,f]$
\eqref{act} with parameters $\{\beta_n\}$ to a similar action with
different parameters $\{\hat\beta_m \sim\beta_{4-n}\}$. In section
\ref{sec:behav}, we consider symmetric models with
$\alpha^{4-n}\beta_n=\alpha^n \beta_{4-n}$, when \eqref{intsym} maps
the model back to itself.

\subsection{Proportional backgrounds} 
Many of the results in the present paper involve proportional
backgrounds, but the implications extend to the general theory. A
useful class of solutions to the bimetric equations \eqref{gf-eq} is
given by the ansatz $\bfmn=c^2\bgmn$. This reduces \eqref{gf-eq} to
two copies of Einstein equations \cite{Hassan:2012wr},     
\beqn
&\bar\G_{\mu\nu}=-\Lambda\,\bgmn\,,\,\,\, \qquad
&\Lambda=\tfrac{m^4}{m_g^2}(\beta_0+3c\beta_1+3c^2\beta_2+c^3\beta_3)\\
&\bar{\tilde\G}_{\mu\nu}=-\tilde\Lambda\,c^2\bgmn\,, \qquad 
&\tilde\Lambda=\tfrac{m^4}{m_f^2c^4}(c\beta_1+3c^2\beta_2+3c^3\beta_3
+c^4\beta_4)\,.
\eeqn
Since $\bar\G_{\mu\nu}=\bar{\tilde\G}_{\mu\nu}$, the above equations
imply,
\be
\Lambda=c^2\tilde\Lambda\,.\label{c}
\ee 
This is a quartic equation that determines $c$ in terms of
the parameters $\beta_n$ and $\alpha$. Obviously $\bgmn$ and $\bfmn$
now coincide with solutions in GR with appropriate cosmological
constants. Generic matter couplings of $g$ and $f$ drive the
solutions away from proportional backgrounds.

These solutions are useful in elucidating the behavior of bimetric
theories. Only around such backgrounds the theory has a well defined
mass spectrum where fluctuations $\delta g=g-\bar g$ and $\delta
f=f-\bar f$ decompose into a massless mode $\delta g+\alpha^2\delta f$
and a massive mode $c^2\delta g-\delta f$ of constant Fierz-Pauli mass
\cite{Hassan:2012wr},  
\beqn\label{FPm}
m_\mathrm{FP}^2=m_0^2(c\beta_1+2c^2\beta_2+c^3\beta_3)\,.
\eeqn
$m_0^2$ depends on normalization of the fluctuations. The presence of
the massive fluctuation distinguishes the bimetric background from the
corresponding GR solution. Some particular backgrounds involving a
mass scale $M_b$ are rendered unstable by the massive mode for,  
\be 
0<m_\mathrm{FP}< M_b\,.
\ee 
For dS backgrounds $M_b^2=2\Lambda/3$ \cite{Higuchi:1986py} and for the
Schwarzschild solutions, $M_b\sim 1/r_h$, where $r_h$ is the black hole 
radius \cite{Babichev:2013una,Brito:2013wya}. For such solutions we restrict to
$m_\mathrm{FP}> M_b$. When $m^2_\mathrm{FP}= 2\Lambda/3$, the linear
theory develops an extra ``partially massless'' symmetry
\cite{Deser:2001us} and propagates 6, rather than 7, modes
\cite{Hassan:2012gz}.

\subsection{The massive gravity limit}\label{sec:mg}
Here we review the massive gravity limit of bimetric theory,
emphasizing a caveat. Massive gravity, with $\gmn$ as the massive
field, is obtained in the limit \cite{Baccetti:2012bk,Hassan:2011zd},  
\beqn
\alpha\longrightarrow\infty\,, \qquad m_g=\mathrm{fixed}\,,\qquad
\alpha^{-2}\beta_4=\beta'_4=\mathrm{fixed}\,, 
\label{mglim}
\eeqn
while $\beta_n$, for $n<4$, are required to remain finite. The scaling 
of $\beta_4$ allows for massive gravity with non-flat reference
metrics. This limit can be taken in two ways:  

1) Without reference to specific solutions, one can take the limit
\eqref{mglim} on the bimetric equations \eqref{gf-eq} (not on
the action that contains $m_f^2\rightarrow\infty$),  
\begin{align}
&\G_{\mu\nu}+m_g^{-2} m^4 V_{\mu\nu} =0\,,\label{mggeq}\\
&\tilde\G_{\mu\nu}+\lambda_f \fmn =0\,,
\qquad \lambda_f=\tfrac{\beta'_4 m^4}{m_g^2}\,,.
\label{mgfeq}
\end{align}
The first equation follows from the massive gravity action for a
general reference metric $\fmn$ \cite{Hassan:2011vm,Hassan:2011tf} (or
from \cite{deRham:2010kj} for $\fmn=\eta_{\mu\nu}$). The second equation
requires that $\fmn$ solves the vacuum Einstein equation, but a
specific solution needs to be selected by hand. It is obvious 
that both of these equations cannot be obtained from a single action.

2) The problem with the absence of an action and the arbitrariness of
$\fmn$ can be circumvented by taking the massive gravity limit in the
bimetric action, after expanding it around a bimetric solution
\cite{Fasiello:2013woa, deRham:2014zqa}: Consider a solution (or family of
solutions) $\fmn^0(\alpha)$ in bimetric theory and expand the action
\eqref{act} around it for canonically normalized perturbations
$\chi$,
\be
\fmn = \fmn^0(\alpha)+m_f^{-1}\chi_{\mu\nu }\,.
\ee
Since the new action will not contain positive powers of $m_f$, the
limit \eqref{mglim} can be taken to obtain a massive gravity action
with a specific reference metric $\fmn^0(\alpha=\infty)$. Equation
\eqref{mgfeq} becomes redundant and is replaced by a decoupled linear
field $\chi_{\mu\nu}$.\footnote{The appearance of linear fields in a
  nonlinear action is consistent only in the absence of general
  covariance.} This way of taking the limit is cumbersome in practice,
but it justifies regarding massive gravity as a limit of bimetric
theory (for a specific bimetric solution).  

In the above construction, a crucial point to note is that {\it
  massive gravity limits exist only for bimetric solutions that remain
  non-singular in the limit $\alpha\rightarrow \infty$}. Then the
limit further truncates the space of such solutions down to the
subclass that reduces to the specified $\fmn^0(\alpha=\infty)$ ({\it
  e.g.}, choosing de Sitter over Schwarzschild-de Sitter). 

In section \ref{sec:behav}, we will consider bimetric models with
solutions that become singular and do not admit a massive gravity
limit. We also discuss the implication of this for the massive gravity
limit of the candidate ``partially masslessness'' bimetric model.

\section{Results on Einstein solutions in bimetric theory}
\label{sec:conseq}
In this section we will show that in any bimetric theory if one of the
metrics is an Einstein metric, the equations force the other one to
be Einstein as well. For certain classes of ghost-free models, that we call the
$\beta_i$-models, the constraint is even stronger: If either metric is
an Einstein metric, then the two metrics have to be proportional to
each other. Finally we point out that the $\beta_i$-models avoid
solutions with determinant singularities and nonlinear ghosts.   

\subsection{General case: Einstein implies Einstein}\label{sec:con}
Using ${(S^2)^\rho}_\sigma =g^{\rho\lambda}f_{\lambda\sigma}$, the
contributions $V_{\mu\nu}$ and $\tilde V_{\mu\nu}$ \eqref{VsDef} to
the bimetric equations become, 
\begin{align}
&g^{\rho\mu}V_{\mu\nu}=V\delta^\rho_\nu-2g^{\rho\mu}
\frac{\p V}{\p {(S^2)^\lambda}_\sigma}\frac{\p
  {(S^2)^\lambda}_\sigma}{\p g^{\mu\nu}}=V\delta^\rho_\nu  
-2\frac{\p V}{\p {(S^2)^\mu}_\sigma}g^{\rho\mu}f_{\sigma\nu}\,,   
\label{gder}\\
&\sqrt{|g^{-1}f|}\,f^{\rho\mu}\tilde{V}_{\mu\nu}=-2f^{\rho\mu}
\frac{\p V}{\delta {(S^2)^\lambda}_\sigma}
\frac{\p {(S^2)^\lambda}_\sigma}{\p f^{\mu\nu}}=
2\frac{\p V}{\p {(S^2)^\mu}_\sigma}g^{\rho\mu}f_{\sigma\nu}\,.
\label{fder}
\end{align}
An immediate consequence is the identity,\footnote{This identify was
  first noticed in \cite{Volkov:2012zb} for the potential $V$ in
  \eqref{vdef}. The derivation here shows that it holds for any
  scalar function of $g^{-1}f$.}    
\beqn\label{potrel}
\sqrt{|g|}\,g^{\rho\sigma}V_{\sigma\nu} + \sqrt{|f|}\,f^{\rho\sigma}
\tilde V_{\sigma\nu}-\sqrt{|g|}\,V\delta^\rho_{~\nu}=0\,.
\eeqn
Using this in the sum of the equations of motion (\ref{gf-eq}) gives, 
\beqn\label{effg}
g^{\rho\sigma}\G_{\sigma\nu}+\alpha^{2}\sqrt{|g^{-1}f|}\,
f^{\rho\sigma}{\tilde\G}_{\sigma\nu}+m_g^{-2}m^4\,V
\delta^\rho_{~\nu}=0\,. 
\eeqn
Now suppose that one of the two metrics, say $\fmn$, satisfies vacuum 
Einstein equations with cosmological constant $\tilde\Lambda$, i.e.~${\tilde\G}_{\mu\nu}+\tilde\Lambda\fmn=0$. Then,
\beqn
g^{\rho\sigma}\G_{\sigma\nu}=\left(\alpha^{2}\tilde\Lambda\sqrt{|g^{-1}f|} 
-m_g^{-2}m^4 V\right)\delta^\rho_{~\nu}\,.
\eeqn
and the Bianchi identity, $\nabla_\rho{\G}^\rho_{~\nu}=0$ gives,
\beqn
\p_\nu \left(\alpha^{2}\tilde\Lambda\sqrt{|g^{-1}f|}-m_g^{-2}m^4
\,V\right)=0\,. 
\eeqn
It is then obvious that $\gmn$ also satisfies $\G_{\mu\nu}+\Lambda 
\gmn=0$ with a constant $\Lambda=m_g^{-2}m^4\,V-\alpha^{2}
\tilde\Lambda\det S$. Of course, this argument also works starting
with $\gmn$ as an Einstein metric. 

This shows that {\it in any bimetric theory, if either one of the two
  metrics is Einstein, then the equations of motion force the other
  one to be Einstein as well}. The proportional solutions,
$\fmn=c^2\gmn$, discussed in the previous section are special cases of
such Einstein backgrounds. But generically, the two Einstein metrics
are not necessarily proportional (due to the polynomial matrix
structure of $V$).

In contrast to the bimetric result, in massive gravity the reference
metric is always Einstein but there is no such constraint on the
second metric. The constraint is lost in the massive gravity limit. In
this limit, a number of different bimetric solutions for $\fmn$ reduce
to the same Einstein metric, while the corresponding solutions for
$\gmn$ remain distinct. Thus, this correspondence between
$\gmn$ and $\fmn$ solutions is lost in the limit.

\subsection*{The general relativity limit:}
When $\alpha\rightarrow 0$, for fixed $m_g$ (i.e.~small $m_f$), the
fluctuation $\delta g$ around proportional backgrounds becomes
massless. This is the General Relativity limit
\cite{Hassan:2011zd,Hassan:2012wr}. The identity \eqref{potrel}
implements this nonlinearly. In this limit, the $\fmn$ equation is
$\tilde V_{\mu\nu}=0$ and \eqref{potrel} implies that $\gmn$ satisfies
the Einstein equation with cosmological constant $V=\,$const., as
observed in \cite{Hassan:2011vm,Baccetti:2012bk}. The same holds in
the presence of a conserved $\gmn$ source. Therefore, any bimetric
model has solutions that can be brought arbitrarily close to GR
solutions by adjusting $\alpha$, irrespective of the mass scale,
avoiding the vDVZ discontinuity (this obviously is not possible in
massive gravity, where $\alpha$ has already been taken to be
infinite). Small $m_f$ is of course a strong coupling limit for the
$\fmn$ sector. In the limit the effect of this strong coupling is to
suppress the dynamics of $\fmn$ and completely determine it in terms
of $\gmn$ such that the later becomes a GR solution. The $\gmn$ sector
however does not exhibit strong coupling.
 
\subsection{The $\beta_i$-models: Einstein implies proportional}
The above result holds for any bimetric theory with a covariant
interaction potential. Now we consider a more restricted class of
models where, of the five $\beta_n$ parameters ($n=0,\cdots,4$) in
\eqref{vdef}, the non-zero ones are $\beta_0$ $\beta_4$, and only one
of the remaining $\beta_i$, where $i=1,2,3$. Following
\cite{Konnig:2013gxa}, we call these the $\beta_i$-models. For these
models we show that if one metric is Einstein, then the other metric
is not only Einstein, but also proportional to the first metric.

\subsection*{The $\beta_1$ and $\beta_3$-models}
Let us first consider the $\beta_1$-model (where $\beta_2=\beta_3=0$). 
In this case the $\gmn$ equations is, 
\beqn\label{geqsm}
g^{\rho\sigma}\G_{\sigma\nu}+\tfrac{m^4}{m_g^2}\Big(\beta_0\delta^\rho_{~\nu} 
+\beta_1\left[S^\rho_{~\nu}-e_1(S)\delta^\rho_{~\nu}\right]\Big)
=0\,. 
\eeqn
Now assume that $\gmn$ is an Einstein metric, $\G_{\sigma\nu}=
-\Lambda g_{\sigma\nu}$. Since $e_1(S)=\Tr \,S$, it is easy to see
that (\ref{geqsm}) implies $S^\rho_{~\nu}=c\delta^\rho_{~\nu}$ with
$c=\frac{1}{3\beta_1}\left(\beta_0-\frac{m_g^2\Lambda}{m^4}\right)$.
Since $S^2=g^{-1}f$ it follows that $\fmn=c^2\gmn$. Conversely, as
demonstrated in the previous subsection, assuming $\fmn$ to be
Einstein always forces $\gmn$ to be Einstein as well and again the
above argument applies.

Due to the interchange property of the potential (\ref{eid}), the
same conclusions can be drawn for the $\beta_3$-model (defined by
$\beta_1=\beta_2=0$), now using the $\fmn$ equation for $S^{-1}$. This
shows that, in the $\beta_1$ and $\beta_3$-models, if either metric
is assumed to be Einstein, then the equations force the other metric
to be not only Einstein, but of the form $f=c^2 g$.

\subsection*{The $\beta_2$-models}\label{sec:b1b3}
Now consider the $\beta_2$-models (defined by $\beta_1=\beta_3=0$)
and assume that one of the metrics is Einstein. Then, by the general
arguments in section \ref{sec:con}, the second metric will also be
Einstein. Again we denote the respective cosmological constants by
$\Lambda$ and $\tilde\Lambda$. In terms of the constants
$\lambda_g\equiv\beta_0-m_g^2m^{-4}\Lambda$ and 
$\lambda_f\equiv\beta_4-m_f^2m^{-4}\tilde\Lambda$, the
equations of motion become, 
\beqn
\lambda_g\delta^\rho_{~\nu}+\beta_2\left[S^\rho_{~\sigma}S^\sigma_{~\nu}
-e_1(S)S^\rho_{~\nu}+e_2(S)\delta^\rho_{~\nu}\right]&=&0\,,
 \label{geqds}\\
\lambda_f\delta^\rho_{~\nu}+\beta_2\left[(S^{-1})^\rho_{~\sigma}
(S^{-1})^\sigma_{~\nu}-\frac{e_3(S)}{e_4(S)}(S^{-1})^\rho_{~\nu}+
\frac{e_2(S)}{e_4(S)}\delta^\rho_{~\nu}\right] &=&0\,.
\label{feqds}
\eeqn
In the second equation, we have used the identity listed above
(\ref{eid}) to re-express $e_n(S^{-1})$ in terms of $e_n(S)$. Tracing
the above equations gives, 
\be
2\lambda_g+\beta_2e_2(S)=0\,,\qquad
2\lambda_fe_4(S)+\beta_2 e_2(S)=0\,.\label{trgf}
\ee
It follows that $e_2(S)$ and $e_4(S)$ are constant,\footnote{
 $\lambda_{f}=0$ or $\lambda_{g}=0$ are ruled out since
  they require $S=0$ which makes the starting equations ill defined.} 
\beqn\label{24const}
e_2(S)=-\frac{2\lambda_g}{\beta_2}\,,\qquad e_4(S)=
\frac{\lambda_g}{\lambda_f}\,.
\eeqn
Let us also multiply the equations (\ref{geqds}) and (\ref{feqds}) by
$S$ and $S^{-1}$, respectively, and take the trace of the
results. Making use of (\ref{eid}) and (\ref{Yid}) we then obtain, 
\be
\lambda_ge_1(S)+3\beta_2e_3(S)=0\,,\qquad
\lambda_fe_3(S)+3\beta_2e_1(S)=0\,.
\ee
These equations fix the value of one of the cosmological constants
because they imply,
\beqn\label{constl}
\lambda_g\lambda_f=9\beta_2^2\,.
\eeqn
Moreover, the relations (\ref{trgf}) can be used to bring the equations of
motion into the form,  
\beqn
\lambda_g\delta^\rho_{~\nu}-\beta_2\left[S^\rho_{~\sigma}S^\sigma_{~\nu}
-e_1(S)S^\rho_{~\nu}\right]&=&0\,,
 \label{geqn}\\
\lambda_f\delta^\rho_{~\nu}-\beta_2\left[(S^{-1})^\rho_{~\sigma}
(S^{-1})^\sigma_{~\nu}-\frac{e_3(S)}{e_4(S)}(S^{-1})^\rho_{~\nu}\right]&=&0\,. 
\label{feqn}
\eeqn
Multiplying the second equation by $\beta_2 S^2$ and subtracting it
from $\lambda_f$ times the first gives, 
\beqn
\left[\lambda_f\beta_2e_1(S)+ \beta_2^2\frac{e_3(S)}{e_4(S)}\right]
S^\rho_{~\nu}+\left(\lambda_f\lambda_g-\beta_2^2\right)\delta^\rho_{~\nu}
=0\,.
\eeqn
This result can be simplified further using (\ref{24const}) and
(\ref{constl}) to get,
\beqn\label{hwt}
3\beta_2e_1(S)S^\rho_{~\nu}+4\lambda_g\delta^\rho_{~\nu}=0\,.
\eeqn
Tracing this expression yields,\footnote{Imaginary solutions are not 
  ruled out since the original equations contain only even powers of $S$.}   
\beqn
e_1(S)=\pm 4\sqrt{-\tfrac{\lambda_g}{3\beta_2}}\,.
\eeqn
Plugging this back into (\ref{hwt}), finally leads to,
\beqn
S^\rho_{~\nu}=c\,\delta^\rho_{~\nu}\,, \qquad
c=\pm\sqrt{\frac{m_g^2\Lambda/m^4-\beta_0}{3\beta_2}}\,, 
\eeqn
which implies $\fmn=c^2\gmn$. Hence, also in this model, assuming 
either metric to be a solution to Einstein's vacuum equations
inevitably leads to proportional solutions.   

\subsection{Effect of matter couplings}
Let us briefly comment on how the situation generalizes when one
includes couplings of the two metrics to matter. The equations
of motion with stress-energy tensors $T_{\mu\nu}$ and ${\tilde
  T}_{\mu\nu}$ are,  
\beqn
g^{\rho\sigma}\G_{\sigma\nu}+\tfrac{m^4}{m_g^2}g^{\rho\sigma}V_{\sigma\nu}
&=~\tfrac{1}{m_g^2}g^{\rho\sigma}T_{\sigma\nu}\,,\label{geqm}\\ 
\alpha^2f^{\rho\sigma}{\tilde\G}_{\sigma\nu}+\tfrac{m^4}{m_g^2}f^{\rho\sigma}
\tilde V_{\sigma\nu}&=~\tfrac{1}{m_g^2}f^{\rho\sigma}
{\tilde T}_{\sigma\nu}\,.
\label{feqm}  
\eeqn
The Bianchi constraints only force a combination of the sources to be
conserved. But demanding the separate general covariance of the matter
couplings in the action implies 
$\nabla^\rho T_{\rho\nu}=\tilde{\nabla}^\rho{\tilde T}_{\rho\nu}=0$, which
is what we assume. While without sources, Einstein solutions always 
exist, in the presence of sources, GR solutions exist only when
allowed by the source. When this is the case, following the same
strategy as before, it is easy to see that the identity (\ref{potrel})
together with the equations of motion now imply the following: 
\vspace{-2mm}
\begin{itemize} \itemsep0em  
\item If one of the metrics solves Einstein's equations with a source,
  i.e., it is a GR solution, then the other metric will also be a GR
  solution, for appropriate sources. A few non-proportional GR
  solutions with sources have been found, see e.g.,
  \cite{Gratia:2013uza}.   
\item Restricting to the $\beta_i$-models will again force the GR
  metrics to be proportional, $\fmn=c^2\gmn$, provided the sources
  satisfy $\alpha^2T_{\mu\nu}={\tilde T}_{\mu\nu}$. Unless the sources
  meet this condition on-shell, there exist no GR solutions for
  either one of the metrics.
\end{itemize}
\vspace{-2mm}
\subsection{Implications for ill-behaved solutions}
Classical solutions of the bimetric theory have been
obtained for cosmological as well as spherically symmetric metrics
(see e.g.~\cite{Volkov:2011an,vonStrauss:2011mq,Comelli:2011zm,
Comelli:2011wq,Volkov:2012wp}).\footnote{For further work of cosmological 
relevance see e.g.~\cite{Akrami:2012vf,Solomon:2014dua,Tamanini:2013xia}} 
Along with good solutions, sometimes
one also finds another branch of solutions with pathologies, in common
with massive gravity. Examples are:
\begin{romanlist}[(ii)] \itemsep0em  
\item Solutions with determinant singularities where the ratio
  $\det{f}/\det{g}$ becomes singular, while curvature invariants are
  well behaved \cite{Gratia:2013uza}. 
\item Solutions with nonlinear ghosts where the
  linear and nonlinear theories have different numbers of degrees of
  freedom \cite{Gumrukcuoglu:2011zh, DeFelice:2012mx, Comelli:2014bqa}.  
\end{romanlist}
\vspace{-2.5mm}
These pathological branches always include solutions where one of the 
metrics solves Einstein's equations while the other metric is not
proportional to it. It is then obvious that such pathological
solutions cannot arise in the $\beta_i$-models, where Einstein
backgrounds are inevitably proportional backgrounds, as shown in the
previous section. 

Let us comment on this advantage of the $\beta_i$ models in some more
detail in the context of cosmological solutions. For homogeneous and
isotropic metrics (see e.g.~\cite{vonStrauss:2011mq}), the Bianchi
constraint~(\ref{bianchi}) factorizes in a way that allows for two
disconnected branches of solutions,  
\beqn
\label{biaf}
(\beta_1\Upsilon+2\beta_2\Upsilon^2+\beta_3\Upsilon^3)\,B=0\,.
\eeqn
Here, $\Upsilon$ is the ratio of the spatial scale factors and $B$ is
a function of metric components and their time-derivatives. Clearly,
there are two ways to solve (\ref{biaf}), either by setting the
bracket to zero or by requiring $B=0$. These two options are called 
``branch I" and ``branch~II" solutions, respectively. It is branch I 
that gives rise to the pathologies mentioned above. In the $\beta_i$
models, this branch does not exist since $\Upsilon=0$ results in a 
singular metric.

Note that the bracket which vanishes on the branch~I solutions becomes
proportional to the Fierz-Pauli mass~(\ref{FPm}) when $\Upsilon$ is
identified with the constant $c$ of the proportional backgrounds. One
can even define an effective (time-dependent) mass for the scalar
perturbations which vanishes on branch~I~\cite{DeFelice:2014nja}. As
discussed in \cite{Gumrukcuoglu:2011zh, DeFelice:2012mx}, this branch
appears to be a pathological class of cosmological solutions.  An
abrupt loss of degrees of freedom occurs in the linearized theory,
accompanied by a non-perturbative, genuine ghost
instability.\footnote{Branch II solutions could also have usual types
  of instabilities in some parameter regions, related to strong
  coupling or violation of the Higuchi bound \cite{Comelli:2012db,
    Comelli:2014bqa}. These are not on the same footing as branch I
  problems.} On top of
that, the authors of~\cite{Gratia:2013uza} found that determinant
singularities are likely to be generic within branch I of the
cosmological backgrounds in bimetric theory. These problem are avoided
by models that do not allow for branch I solutions, which are
precisely the $\beta_i$ models.

A seemingly unrelated question, first raised in
\cite{Baccetti:2012re}, is the violation of the null energy condition
for the effective stress-energies as set up by the interaction
potential. In general it is not completely clear what such a violation
means physically since the bimetric interactions considered here are
not part of the matter sector. It is nevertheless interesting that
\cite{Baccetti:2012re} concluded that one of the effective
stress-energies must always violate the null energy condition apart
from the exceptional case when the metrics are proportional. Within
the $\beta_i$-models, whenever one metric
is Einstein and has a well-defined notion of energy connected to it,
then the two metrics are forced to be proportional and hence avoid any
pathologies connected to a possible violation of the null-energy
condition.

\section{Bimetric solutions with no massive gravity limit}
\label{sec:behav}  
In this section we focus on bimetric solutions that do not admit a
massive gravity limit. As explicit examples, we first consider
proportional solutions and then study the behavior of generic
solutions in a class of ``symmetric'' bimetric models. Finally we
discuss the implications of this for the massive gravity limit of the 
candidate PM bimetric model (which is a symmetric model as well 
as a $\beta_2$-model). 

\subsection{Proportional solutions with no massive gravity limit}
As in section \ref{sec:mg}, we define the massive gravity limit as
$\alpha=m_f/m_g\rightarrow \infty$ at fixed $m_g$, so that $\gmn$
becomes the massive field. As emphasized there, to find a massive
gravity action, one needs bimetric solutions with a good massive
gravity limit, which is not always the case. 

As a warmup, consider proportional solutions $f=c^2g$ in a bimetric
model with only $\beta_0$ and $\beta_1$ nonvanishing. In this case, 
equation \eqref{c} gives two solutions for $c$, 
\be
c_{\pm}=\tfrac{1}{6\beta_1}\left(-\beta_0\pm\sqrt{\beta_0^2+12
  \alpha^{-2}\beta_1^2} \right) 
\ee 
In the massive gravity limit, $c_-=-\beta_0/3\beta_1$ while $c_+=0$
(for $\beta_0>0$). Hence the class of bimetric solutions that are
connected to $c_+$ do not have a continuous massive gravity limit with
both metrics remaining well defined. In this case, formally setting
$\fmn=0$ in the action (ignoring that the curvatures are not defined
and the equations break down) will eliminate the mass potential giving
massless GR which is not a continuous limit of massive gravity. Below
we consider models where both metrics become singular in the limit.

We now turn to an interesting class of ``symmetric" bimetric models with
parameters $\beta_n$ satisfying, 
\be
\label{betaselfdual}
\alpha^{4-n}\beta_n=\alpha^n\beta_{4-n}\,. 
\ee
Then the bimetric action is invariant under the interchange,
\be
\label{intsym2}
\alpha^{-1}\gmn\leftrightarrow\alpha\fmn\,.  
\ee 
without having to interchange the $\beta_n$. In other words, the
interchanges (\ref{intsym}) now map the model to itself.\footnote{In the presence of matter the interchange symmetry can only be preserved for very specific couplings to the metrics. Here we only consider the classical vacuum equations, in particular since the presence of generic matter couplings would anyway spoil the possible nonlinear PM symmetry discussed in a subsequent section. For recent work on the consistency of matter couplings in bimetric theory see~\cite{Yamashita:2014fga,deRham:2014naa}.} 

In order to get a finite massive gravity limit with nontrivial
interactions, we need to further specify the scaling of the parameters
with $\alpha$ by assuming fixed $\beta_2$ and $\beta_3$ and setting
$\beta_4=\alpha^2\beta'_4$ with fixed $\beta'_4$. For this class of
models, the bimetric equations of motion become,  
\begin{align}
\G^\rho_{~\nu}&=-\tfrac{m^4}{m_g^2}\left(\alpha^{-2}\beta_4'\,
\mathbb 1+\alpha^{-2}\beta_{3}Y_{(1)}(S)+
\beta_{2}Y_{(2)}(S)+\beta_{3}Y_{(3)}(S)\right)^\rho_{~\nu}\,, 
\label{eqbsm1}\\
{\tilde\G}^\rho_{~\nu}&=-\tfrac{m^4}{m_g^2}\left(\beta_4'\,
\mathbb 1+\alpha^{-2}\beta_{3}Y_{(1)}(S^{-1})+
\alpha^{-2}\beta_{2}Y_{(2)}(S^{-1})+\alpha^{-4}\beta_{3}
Y_{(3)}(S^{-1})\right)^\rho_{~\nu}\,.
\label{eqbsm2}
\end{align}
The massive gravity limit of the equations can be found easily. The
scalings chosen insure the survival of the $\beta_2$ and $\beta_3$
interactions in the $\alpha\rightarrow\infty$ limit but, at the same
time, $\beta_0 =\alpha^{-2}\beta'_4$ and $\beta_1 =\alpha^{-2}\beta_3$
both disappear. This is the most general scaling behavior of the
parameters \eqref{betaselfdual} with a well-defined massive gravity
limit for the equations. 

Let us now consider proportional solutions $\fmn=c^2\gmn$ to the above
equations. In this case equation \eqref{c} which determines $c$ becomes, 
\be
\alpha^{-1}\beta_3+(3\beta_2-\beta'_4)\alpha c-(3\beta_2-\beta'_4)
\alpha^3 c^3-\beta_3\alpha^3c^4=0\,.
\ee
For general $\beta_2$, $\beta_3$ and $\beta'_4$, this has four
solutions,  
\beqn\label{solc}
c_{(1)}^\pm=\pm\frac{1}{\alpha}\,,\qquad 
c_{(2)}^\pm=\frac{3\beta_2-\beta'_4}{2\beta_3}\left( -1\pm\sqrt{1-
\tfrac{4\beta_3^2}{\alpha^2(3\beta_2-\beta'_4)^2}}\right)\,.
\eeqn
The interchange $\alpha^{-1}\gmn\leftrightarrow\alpha\fmn$ amounts to
$\alpha c\rightarrow \alpha^{-1}c^{-1}$ on the above solutions. While
this keeps the first two solutions in (\ref{solc}) invariant, the
other two transform into each other.

In the limit, $\alpha\rightarrow \infty$, only $c_{(2)}^-$ gives rise
to a well defined massive gravity, 
\be
c_{(2)}^-\rightarrow \frac{\beta'_4-3\beta_2}{\beta_3}\,.
\ee
The other three solutions vanish and at least one of the
metrics becomes singular. In particular, the solution $c_{(2)}^+$
obtainable from $c_{(2)}^-$ through the interchange \eqref{intsym2}
becomes ill-defined. Classes of bimetric solutions connected to these
are lost in the massive gravity limit. 

The class of models considered above contains the candidate PM
bimetric model, discussed in section \ref{sec:PM}. A motivation for
showing the existence of solutions with no massive gravity limit is to
argue that such a limit of the candidate PM bimetric theory involves a
truncation of the solution space that also destroys a possible
on-shell gauge symmetry of the theory, except around very special
backgrounds. But, considering proportional backgrounds alone is not
enough for this argument, since in the PM case equation \eqref{c}
leaves $c$ undetermined. In that case, $c$ and its $\alpha$-dependence
are gauge dependent. To overcome this, in the following we consider the
massive gravity limit of generic solutions.

\subsection{Generic solutions with a massive gravity limit} 

For the bimetric equations \eqref{eqbsm1}-\eqref{eqbsm2}, we first 
consider the general form of solutions with a good massive gravity 
limit. Solutions without such a limit are obtained in the next
subsection. Arranged in powers of $\alpha$, equations
\eqref{eqbsm1}-\eqref{eqbsm2} have the form, 
\beqn
\G(g)&=&V^{(0)}(g,f)+\alpha^{-2}V^{(2)}(g,f)\,,\label{oeq1}\\
{\tilde\G}(f)&=&\tilde{V}^{(0)}(f)+\alpha^{-2}\tilde{V}^{(2)}(g,f)+
\alpha^{-4}\tilde{V}^{(4)}(g,f)\,,  
\label{oeq2}
\eeqn
where $\G$ is the matrix notation for the Einstein tensor
$\G_{\mu\nu}$ and the matrices $V^{(n)}$ and $\tilde{V}^{(n)}$ can be
read off from \eqref{eqbsm1}-\eqref{eqbsm2} by comparing powers of
$\alpha$. 

In the limit $\alpha\rightarrow \infty$, only the lowest-order,
$\alpha$-independent terms survive. Hence, for large enough $\alpha$,
all bimetric solutions with a massive gravity limit must be expandable 
as,     
\beqn\label{ans}
\fmn=\bar{f}_{\mu\nu}+\alpha^{-2}\psi_{\mu\nu} +\mathcal{O}
(\alpha^{-4})\,,\qquad 
\gmn=\phi_{\mu\nu}+\alpha^{-2}\chi_{\mu\nu}+\mathcal{O}(\alpha^{-4})\,. 
\eeqn
Since these solve the equations to order $\alpha^{-2}$, the above
functions must satisfy,  
\begin{align}
\G(\phi)&=V^{(0)}(\phi,\bar{f})\,,\qquad\left.\frac{\delta(\G+V^{(0)})}
{\delta \gmn}\right|_{\bar{f},\, \phi}\chi_{\mu\nu}+\left.
\frac{\delta V^{(0)}}{\delta\fmn}\right|_{\bar{f},\,\phi}\psi_{\mu\nu} 
+V^{(2)}(\phi,\bar{f})=0\,,  \label{soleqp1}\\
{\tilde\G}(\bar{f})&=\tilde{V}^{(0)}(\bar{f})\,, \qquad \quad 
\left.\frac{\delta({\tilde\G}+\tilde{V}^{(0)})}{\delta\fmn}
\right|_{\bar{f}}\psi_{\mu\nu}+\tilde{V}^{(2)}(\phi,\bar{f})=0\,.
\label{soleqp2}
\end{align}
Note that since ${\tilde\G}(\bar{f})=\tilde{V}^{(0)}(\bar{f})$ is
simply an Einstein equation with cosmological constant, this implies
that $\bfmn$ is Einstein. Accordingly, $\left.\frac{\delta({\tilde\G}
+\tilde{V}^{(0)})}{\delta\fmn}\right|_{\bar{f}}$ is the linearized
Einstein operator. The lowest order term $\phi_{\mu\nu}$ in the
solution for $\gmn$, as well as the $\mathcal{O}(\alpha^{-2})$
contributions $\psi_{\mu\nu}$ and $\chi_{\mu\nu}$ in both metrics are
generic ($\alpha$-independent) functions determined by the above
equations. 

Solutions of the form (\ref{ans}) must exist in bimetric theory, or
else there would be no massive gravity limit. In the limit,
$\fmn=\bfmn$ is Einstein and $\gmn=\phi_{\mu\nu}$ possibly differs
from an Einstein metric. This also explicitly captures the constraint
on bimetric solutions that is lost in massive gravity: If $\gmn$ in
bimetric theory is not Einstein then neither is $\fmn$, but its difference
from the Einstein metric $\bar{f}_{\mu\nu}$ vanishes in the massive
gravity limit, while $\gmn$ can remain a non-Einstein metric.

\subsection{Generic solutions without a massive gravity
  limit}\label{sec:singsol} 

The symmetry of equations \eqref{eqbsm1}-\eqref{eqbsm2} under 
(\ref{intsym2}) is explicit in terms of the rescaled variables, 
\beqn
\gmnp\equiv \alpha^{-1} \gmn\,,\qquad \fmnp\equiv\alpha\fmn\,.
\eeqn
In terms of these the equations of motion
\eqref{eqbsm1}-\eqref{eqbsm2} have the form, 
\beqn
\G(g')&=&\alpha^{-1}V^{(0)}(g',f')+\alpha^{-2}V^{(2)}(g',f')\,,\\
{\G}(f')&=&\alpha^{-1}V^{(0)}(f',g')+\alpha^{-2}V^{(2)}(f',g')\,,
\eeqn
which are now manifestly invariant under the interchange,\footnote{The action \eqref{act} also takes a symmetric form, $m_g^2
\alpha\int\dd^4 x\big\{\sqrt{|g'|}[R(g')-2\tilde m^2~(\beta_0+
\beta_1\alpha^{-1} e_1(S')+ \tfrac{1}{2}\beta_2\alpha^{-2}e_2(S'))]  
+\sqrt{|f'|}[R(f')-2\tilde m^2(\beta_0+\beta_1\alpha^{-1}
e_1(S'^{-1})+\tfrac{1}{2}\beta_2\alpha^{-2}e_2(S'^{-1}))]\big\}
$. Although our considerations are completely at the classical level,
quantum corrections are expected to preserve the interchange symmetry
if the quantization procedure respects the symmetry.} 
\beqn\label{symtrres}
\gmnp\longleftrightarrow\fmnp\,.
\eeqn
The solution \eqref{ans} in terms of the new variables reads,
\beqn\label{ansres}
\fmnp=\alpha\left(\bar{f}_{\mu\nu}+\alpha^{-2}\psi_{\mu\nu}\right) +
\mathcal{O}(\alpha^{-3})\,,\qquad 
\gmnp=\alpha^{-1}\left(\phi_{\mu\nu}+\alpha^{-2}\chi_{\mu\nu}\right)+
\mathcal{O}(\alpha^{-4})\,.
\eeqn
The crucial point is that, due to the interchange symmetry, the
above bimetric equations also admit another solution that is obtained
by applying \eqref{symtrres} to \eqref{ansres}. The new solution is, 
\beqn\label{ansresint}
\gmnp=\alpha\left(\bar{f}_{\mu\nu}+\alpha^{-2}\psi_{\mu\nu}\right) 
+\mathcal{O}(\alpha^{-3})\,,\qquad\fmnp=\alpha^{-1}\left(
\phi_{\mu\nu}+\alpha^{-2}\chi_{\mu\nu}\right)+\mathcal{O}(\alpha^{-4})\,,
\eeqn
In terms of the original variables, this becomes,
\beqn\label{ans2}
\gmn=\alpha^2\bfmn+\psi_{\mu\nu}+\mathcal{O}(\alpha^{-2})\,,
\qquad \fmn=\frac{1}{\alpha^2}\phi_{\mu\nu}+\frac{1}{\alpha^4}
\chi_{\mu\nu}+\mathcal{O}(\alpha^{-6})\,,
\eeqn
By inserting this into the equations and separating different orders
of $\alpha$, one can explicitly verify that \eqref{ans2} is a
perturbative solution provided \eqref{soleqp1}-\eqref{soleqp2} hold. 

Now, unlike \eqref{ans}, in the limit $\alpha\rightarrow\infty$ both
metrics become singular and the solution has no massive gravity
limit. This result is general and also holds for the parameter values
corresponding to the candidate PM bimetric theory. 

\subsection{Implications for a nonlinear PM symmetry}\label{sec:PM}

The linear partially massless spin-2 theory is formulated
around de Sitter or Einstein backgrounds~\cite{Deser:2001us}. The
bimetric model with parameters,\footnote{Note that this model is in
  the intersection of the $\beta_i$ models discussed in section
  \ref{sec:conseq} and the symmetric models considered in this
  section. Hence both results hold for this model.}  
\be
\beta_1=\beta_3=0\,,\qquad \alpha^2\beta_0=3\beta_2=\alpha^{-2}\beta_4  
\label{PM}
\ee 
emerges as the unique candidate with the potential to provide a
nonlinear realization of the PM phenomenon in the bimetric framework
\cite{Hassan:2012gz, Hassan:2012rq, Hassan:2013pca}. If so, this
theory must have a gauge invariance and propagate six (rather than
seven) modes. This has been difficult to prove, due to the intricate
structure of the theory, but there is some evidence for it beyond the
linear theory around Einstein backgrounds \cite{Hassan:2013pca,
  prepar}. Hence there is a possibility that this model, or some
appropriate generalization of it, could have a gauge symmetry. On the
other hand, there have also been arguments that such a nonlinear
generalization does not exist \cite{deRham:2013wv, Deser:2013uy,
  Fasiello:2013woa,Joung:2014aba}.\footnote {A detailed analysis of
  these arguments will be presented in \cite{prepar}. There, more
  evidence for the presence of an on-shell gauge symmetry is provided
  and it is also shown that the counter arguments so far do not rule
  out a bimetric theory with a nonlinear symmetry.} Here we focus on
the most common argument, based on the massive gravity limit.

PM massive gravity was investigated in \cite{deRham:2012kf,
  deRham:2013wv} and can be obtained from the bimetric model
\eqref{PM} in the limit $\alpha\rightarrow \infty$ with $\beta_2$ held
fixed and $\beta_0\rightarrow 0$.\footnote{Even non-PM bimetric models
  with $\beta_0=\beta_0'/\alpha^2$ lead to the PM massive
  gravity. Hence a massive gravity background does not have a unique
  bimetric extension.} It has been argued that this limit cannot
exhibit a PM symmetry nonlinearly \cite{deRham:2013wv,
  Deser:2013uy}. It is then stated that since massive gravity is a
well-behaved limit of bimetric theory, the same conclusion must also
apply to the bimetric model \eqref{PM} \cite{deRham:2013wv,
  Deser:2013uy, deRham:2014zqa, Fasiello:2013woa}. This argument
assumes that if a PM bimetric model exists, its gauge symmetry would
survive in the massive gravity limit.

From the discussion above it is clear that massive gravity does not
only involve taking a limit of bimetric theory, but it also amounts to
truncating the space of bimetric solutions to a single class 
leading to a specific massive gravity reference metric. In particular,
bimetric solutions that become singular in the limit are discarded. It
is obvious that if the bimetric model \eqref{PM} has a gauge symmetry
that cannot be truncated in the same way, then it will not survive in
the massive gravity limit. 

There is evidence that this is indeed the case. In
\cite{Hassan:2013pca} it was shown 
that, in a perturbative treatment in powers of curvatures, one can
eliminate one of the metrics, say $\fmn$, between the bimetric
equations \eqref{eqbsm1}-\eqref{eqbsm2} to get a single equation for
$\gmn$. At the lowest order, this gives the equation of motion for
conformal gravity with Weyl scalings as a gauge symmetry. Derivative
corrections to the Weyl scaling can be computed at low orders
\cite{Hassan:2013pca, prepar}. This provides an on-shell, nonlinear and 
background independent extension of the linear PM symmetry.

It turns out that for generic non-proportional solutions, these gauge
transformations diverge in the limit $\alpha\rightarrow \infty$ (when
the perturbative relation to conformal gravity also breaks down). Only for
proportional backgrounds (and simple solutions like the cosmological
ones \cite{vonStrauss:2011mq}), the terms with higher powers of
$\alpha$ in the perturbative expansions vanish and the transformations
admit a massive gravity limit. Thus away from this limited class of
solutions, the known gauge transformations do not admit a massive
gravity truncation. The picture that emerges, then, is that gauge
transformations without a massive gravity limit relate bimetric
solutions with a massive gravity limit to solutions that are singular in
the limit. The existence of such solutions has been explicitly shown
here. Hence we expect that if the bimetric theory \eqref{PM} has a
nonlinear gauge symmetry, this will not survive in the massive gravity
limit, except around a limited set of backgrounds. In particular, they
will not survive for massive gravity backgrounds that appreciably
differ from proportional backgrounds.

\section*{Acknowledgments}

We thank M.~Crisostomi, J.~Enander,
A.~E.~Gumrukcuoglu, W.~Hu, E.~M\"ortsell and L.~Pilo for useful
discussions.  MvS wishes to acknowledge the creative
environment supplied by the SW8 workshop, ``Hot Topics in Modern
Cosmology'', held at IESC in Carg\`{e}se this year. The research of MvS
leading to these results has received funding from the European
Research Council under the European Community’s Seventh Framework
Programme (FP7/2007-2013 Grant Agreement no. 307934).

%
%
%
%


\end{document}